\documentclass[12pt]{article}
\usepackage{graphicx}
\usepackage{amsmath}

\oddsidemargin  - 5 mm
\evensidemargin - 5 mm
\newtheorem{theorem}{Theorem}

\newtheorem{proposition}[theorem]{Proposition}

\input{tcilatex}

\begin{document}

\date{}
\title{Elliptic Curves, Algebraic Geometry Approach in Gravity Theory and 
Uniformization of Multivariable Cubic Algebraic Equations}
\author{Bogdan G. Dimitrov \thanks{%
Electronic mail: bogdan@theor.jinr.ru} \\
%EndAName
Bogoliubov Laboratory of Theoretical Physics\\
Joint Institute for Nuclear Research \\
6 Joliot-Curie str. \\
Dubna 141 980, Russia}
\maketitle

\begin{abstract}
\ \ 

\ \ \ Based on the distinction between the covariant and contravariant
metric tensor components in the framework of the affine geometry approach
and the s.c. "gravitational theories with covariant and contravariant
connection and metrics", it is shown that a wide variety of third, fourth,
fifth, seventh, tenth- degree algebraic equations exists in gravity theory.
This is important \ in view of finding new solutions of the Einstein's
equations, if they are treated as algebraic ones. Since the obtained cubic
algebraic equations are multivariable, the standard algebraic geometry
approach for parametrization of two-dimensional cubic equations with the
elliptic Weierstrass function cannot be applied. Nevertheless, for a
previously considered cubic equation for reparametrization invariance of the
gravitational Lagrangian and on the base of a newly introduced notion of
"embedded sequence of cubic algebraic equations", it is demonstrated that in
the multivariable case such a parametrization is also possible, but with
complicated irrational and non-elliptic functions. After finding the
solutions of a system of first - order nonlinear differential equations,
these parametrization functions can be considered also as uniformization
ones (depending only on the complex uniformization variable $z$) for the
initial multivariable cubic equation.  
\end{abstract}

\section{\protect\bigskip \protect\bigskip INTRODUCTION}

Inhomogeneous cosmological models have been intensively studied in the past
in reference to colliding gravitational \bigskip waves [1] or singularity
structure and generalizations of the Bondi - Tolman and Eardley-Liang-Sachs
metrics [2, 3]. In these models the inhomogeneous metric is assumed to be of
the form [2] 
\begin{equation}
ds^{2}=dt^{2}-e^{2\alpha (t,r,y,z)}dr^{2}-e^{2\beta (t,r,y,z)}(dy^{2}+dz^{2})
\tag{1.1}
\end{equation}%
(or with $r\rightarrow z$ and $z\rightarrow x$), which is called the
Szafron-Szekeres metric [4-7]. In [7], after an integration of one of the
components - $G_{1}^{0}$ of the Einstein's equations, a solution in terms of
an elliptic function is obtained.

In different notations, but again in the framework of the Szafron-Szekerez
approach the same integrated in [7] nonlinear differential equation \ 
\begin{equation}
\left( \frac{\partial \Phi }{\partial t}\right) ^{2}=-K(z)+2M(z)\Phi ^{-1}+%
\frac{1}{3}\Lambda \Phi ^{2}  \tag{1.2}
\end{equation}%
was obtained in the paper [8] of Kraniotis and Whitehouse. They make the
useful observation that (1.2) in fact defines a (cubic) algebraic equation
for an elliptic curve, which according to the standard algebraic geometry
prescribtions (see [9] for an elementary, but comprehensive and contemporary
introduction) can be parametrized with the elliptic Weierstrass function 
\begin{equation}
\rho (z)=\frac{1}{z^{2}}+\sum\limits_{\omega }\left[ \frac{1}{(z-\omega )^{2}%
}-\frac{1}{\omega ^{2}}\right]  \tag{1.3}
\end{equation}%
and the summation is over the poles in the complex plane. Two important
problems immediately arise, which so far have remained without an answer%
\textbf{: }

1. The parametrization procedure with the elliptic Weierstrass function in
algebraic geometry is adjusted for cubic algebraic equations with number
coefficients! Unfortunately, equation (1.2) is not of this type, since it
has coefficient functions in front of the variable $\Phi $, which depend on
the complex variable $z$. In view of this, it makes no sense to define
\textquotedblright Weierstrass invariants\textquotedblright\ as 
\begin{equation}
g_{2}=\frac{K^{2}(z)}{12}\text{ \ \ ; \ \ \ }g_{3}=\frac{1}{216}K^{3}(z)-%
\frac{1}{12}\Lambda M^{2}(z)\text{ \ \ ,}  \tag{1.4}
\end{equation}%
since the above functions have to be set up equal to the complex numbers $%
g_{2}$ and $g_{3}$ (the s. c. Eisenstein series) 
\begin{equation}
g_{2}=60\sum\limits_{\omega \subset \Gamma }\frac{1}{\omega ^{4}}%
=\sum\limits_{n,m}\frac{1}{(n+m\tau )^{4}}\text{ \ \ \ ,}  \tag{1.5}
\end{equation}%
\begin{equation}
g_{3}=140\sum\limits_{\omega \subset \Gamma }\frac{1}{\omega ^{6}}%
=\sum\limits_{n,m}\frac{1}{(n+m\tau )^{6}}\text{ \ \ \ \ }  \tag{1.6}
\end{equation}%
and therefore additional equations\textbf{\ }have to be satisfied in order
to ensure the parametrization with the Weierstrass function.

2. Is the Szekerez - Szafron metric the only case, when the parametrization
with the Weierstrass function is possible? Closely related to this problem
is the following one - is only one of the components of the Einstein's
equation parametrizable with \ $\rho (z)$ and its derivative?

This  paper has the aim to present an adequate mathematical algorithm for
finding solutions of the Einstein's equations in terms of elliptic
functions. This approach is based on the clear distinction between covariant
and contravariant metric tensor components within the s.c affine geometry
approach, which will be clarified further in Section 2. Afterwords, a cubic
algebraic equation in terms of the contravariant metric components will be
obtained, which according to the general prescription and the algorithm in
the previous paper [10] can be parametrized with the Weierstrass function
and its derivative. Respectively, if the contravariant components are
assumed to be known, then a cubic (or a quartic) algebraic equation with
respect to the covariant components can be investigated and parametrized
again with the Weierstrass function. \textit{Thus it will turn out that the
parametrization with the Weierstrass function will be possible not only in
the Szafron-Szekeres case, but also in the general case due to the "cubic"
algebraic structure of the gravitational Lagrangian.} This is an important
point since valuable cosmological characteristics for observational
cosmology such as the Hubble's constant $H(t)=\frac{\overset{.}{R}(t)}{R(t)}$
and the deceleration parameter $q=-\frac{\overset{..}{R}(t)R(t)}{\overset{.}{%
R}^{2}(t)}$ may be expressed in terms of the Jacobi's theta function and of
the Weierstrass elliptic function respectively [8]. Unfortunately, in the
paper [8] the Eisenstein series (1.5-1.6) have not been taken into account,
due to which the obtained expression for the metric will be another one and
will be modified.

Instead of searching out eliptic solutions of the Einstein's equations for
each separate case of a given metric, as in nearly all of the mentioned
papers, in the this paper another method will be proposed. First, a cubic
algebraic equataion will be parametrized with respect to one of the
contravariant components, following the approach in a previous paper [10].
Then, this parametrization will be extended to more than one variable in the 
\textit{multivariable cubic algebraic equation (section 6).} This will be a
substantial and new development, different from the standard algebraic
geometry approach, in which \textit{only two-dimensional cubic equations }%
are parametrized with the (elliptic) Weierstrass function and its
derivative. Finally, the dependence of the generalized coordinates $%
X^{i}=X^{i}(x_{1},x_{2},x_{3},.....,x_{n})$ on the complex variable $z$ will
be established from a derived system of first-order nonlinear differential
equations (section 7). The generalized coordinates can be regarded as $n-$
dimensional hypersurfaces, defining a transition from an initially defined
set of coordinates $x_{1},x_{2},x_{3},.....,x_{n}$ on a chosen manifold to
another set of the generalized coordinates $X^{1},X^{2},.....,X^{n}$. Since
the covariant metric components $g_{ij}$ also depend on these coordinates,
this means that their dependence on the complex variable $z$ will also be
known. In other words, at the end of the applied approach, each initially
given function $g_{ij}(t,\mathbf{x})$ of the time and space coordinates will
be expressed also as $g_{ij}(z)$. The algebraic approach will be applied to
the s .c. \textit{cubic algebraic equation for reparametrization invariance
of the gravitational Lagrangian}, but further it will be shown that not only
the approach will be applicable in the general case of an arbitrary
contravariant tensor, but also concrete solutions for the metric $g_{ij}(z)$
will be given in the case of specially chosen simple metrics.

The present paper continues and develops further the approach from a
previous paper [10], where a definite choice of the contravariant metric
tensor was made in the form of the factorized product $\widetilde{g}%
^{ij}=dX^{i}dX^{j}$. The differentials $dX^{i}$ are assumed to lie in the
tangent space $T_{X}$ of the generalized coordinates. In Section 2 of the
present paper some basic facts about the affine geometry approach and the
s.c. \textit{gravitational theory with covariant and contravariant metrics
and connections (GTCCMC)} will be reminded, which has been described in the
review article [13]. In its essence, the distinction between covariant and
contravariant components is related to the \textit{affine geometry approach}
[15, 16], according to which the four-velocity tangent vector at each point
of the observer's worldline is \textit{not normalized} and \textit{not equal
to one}, i.e. $l_{a}l^{a}=l^{2}\neq 1$. Similarly, for a second-rank tensor
one would have $g_{\mu \nu }g^{\nu \alpha }=l_{\mu }^{\alpha }$ $\neq \delta
_{\mu }^{\alpha }$. In the next section 3 it will be demonstrated briefly
how the cubic algebraic equation with respect to the differentials $dX^{i}$
was derived in [10], but in fact the aim will be to show that depending on
the choice of variables in the gravitational Lagrangian or in the Einstein's
equations, a \textit{wide variety of algebraic equations} (of third, fourth,
fifth, seventh degree) in gravity theory may be treated, if a \textit{%
distinction between the covariant metric tensor components and the
contravariant ones is made.} This idea, originally set up by Schouten and
Schmutzer, was further developed in the papers [13, 14]. In usual gravity
theory, the contravariant components are at the same time inverse to the
covariant ones , and thus the correspondence between \textquotedblright
covectors\textquotedblright\ (in our terminology - these are the
\textquotedblright vectors\textquotedblright ) and the \textquotedblright
vectors\textquotedblright\ (i.e. the contravariant vectors\textquotedblright
) is being set up, since both these kinds of tensors satisfy the matrix
equation $g_{ij}g^{jk}=\delta _{i}^{k}$. However, within the framework of
affine geometry, such a correspondence is not necessarily to be established
(see again [15-18]) and both tensors have to be treated as \textit{different
mathematical objects,} defined on one and the same manifold. 

The physical idea, which will be exploited in this paper will be: \textit{%
can such a gravitational theory with a more general contravariant tensor
have the same gravitational Lagrangian as in the known gravitational theory
with  contravariant metric tensor components, which are at the same time the
inverse ones to the covariant one?} On the base of such an "equivalence" the
s. c. cubic algebraic equation for reparametrization invariance of the
gravitational Lagrangian was obtained in [10]. The derivation was based also
on the construction of another connection $\widetilde{\Gamma }%
_{kl}^{s}\equiv \frac{1}{2}dX^{i}dX^{s}(g_{ik,l}+g_{il,k}-g_{kl,i})$. It can
be proved that the connection $\widetilde{\Gamma }_{kl}^{s}$ has two very
useful properties: 1. It may have an affine transformation law under a broad
variety of coordinate transformations, which can be found after solving a
system of nonlinear differential equations. 2. $\widetilde{\Gamma }_{kl}^{s}$
is an equiaffine connection, which is a typical notion, introduced in
classical affine geometry [15, 16] and meaning that there exists a volume
element, which is preserved under a parallel displacement of a basic $n-$%
dimensional vector $e\equiv e_{i_{1}i_{2}....i_{n}}$. Equivalently defined, $%
\widetilde{\Gamma }_{kl}^{s}$ is an equiaffine connection [15, 16] if it can
be represented in the form $\widetilde{\Gamma }_{ks}^{s}=\partial _{k}lge$,
where $e$ is a scalar quantity. This notion turns out to be very convenient
and important, since for such types of connections we can use the known
formulae for the Ricci tensor, but with the connection $\widetilde{\Gamma }%
_{kl}^{s}$ instead of the usual Christoffell one $\Gamma _{kl}^{s}$.
Moreover, the Ricci tensor $\widetilde{R}_{ij}$ will again be a symmetric
one, i.e. $\widetilde{R}_{ij}=\widetilde{R}_{ji}=\partial _{k}\widetilde{%
\Gamma }_{ij}^{k}-\partial _{i}\widetilde{\Gamma }_{kj}^{k}+\widetilde{%
\Gamma }_{kl}^{k}\widetilde{\Gamma }_{ij}^{l}-\widetilde{\Gamma }_{ki}^{m}%
\widetilde{\Gamma }_{jm}^{k}$.

\section{\ AFFINE GEOMETRY APPROACH \ AND\ GRAVITATIONAL\ \ THEORIES\ \
WITH\ \ COVARIANT\ \ AND\ CONTRAVARIANT METRICS \ AND \ CONNECTIONS \ \ }

This section has the purpose to review some of the basic aspects of \textit{%
\ GTCCMC}, which would further allow the application of algebraic geometry
and theory of algebraic equations in gravity theory.

It is known in gravity theory that the  metric tensor $g_{ij}$ determines
the space - time geometry, which means that the Christoffell connection 
\begin{equation}
\Gamma _{ik}^{l}\equiv \frac{1}{2}g^{ls}(g_{ks,i}+g_{is,k}-g_{ik,s}) 
\tag{2.1}
\end{equation}%
and the Ricci tensor 
\begin{equation}
R_{ik}=\frac{\partial \Gamma _{ik}^{l}}{\partial x^{l}}-\frac{\partial
\Gamma _{il}^{l}}{\partial x^{k}}+\Gamma _{ik}^{l}\Gamma _{lm}^{m}-\Gamma
_{il}^{m}\Gamma _{km}^{l}\text{ \ \ \ }  \tag{2.2}
\end{equation}%
can be calculated.

It is useful to remember also [23] the s. c. Christoffell connection of the
first kind: 
\begin{equation}
\Gamma _{i;kl}\equiv g_{im}\Gamma _{kl}^{m}=\frac{1}{2}%
(g_{ik,l}+g_{il,k}-g_{kl,i})\text{ \ ,}  \tag{2.3}
\end{equation}%
obtained from the expression for the zero covariant derivative $0=\nabla
_{l}g_{ik}=g_{ik,l}-g_{m(i}\Gamma _{k)l}^{m}$ $.$ By contraction of (2.3)
with another contravariant tensor field $\widetilde{g}^{is}$, one might as
well define \textit{another connection}\textbf{: } 
\begin{equation}
\widetilde{\Gamma }_{kl}^{s}\equiv \widetilde{g}^{is}\Gamma _{i;kl}=%
\widetilde{g}^{is}g_{im}\Gamma _{kl}^{m}=\frac{1}{2}\widetilde{g}%
^{is}(g_{ik,l}+g_{il,k}-g_{kl,i})\text{ \ ,}  \tag{2.4}
\end{equation}%
not consistent with the initial metric $g_{ij}$. Clearly the connection
(2.4) is defined under the assumption that the contravariant metric tensor
components $\widetilde{g}^{is}$ are not to be considered to be the inverse
ones to the covariant components $g_{ij}$ and therefore $\widetilde{g}%
^{is}g_{im}\equiv f_{m}^{s}(\mathbf{x})$.

In fact, the definition $\widetilde{g}^{is}g_{im}\equiv f_{m}^{s}$ turns out
to be inherent to gravitational physics. For example, in the projective
formalism one decomposes the standardly defined metric tensor (with $%
g_{ij}g^{jk}=\delta _{i}^{k}$) as 
\begin{equation}
g_{ij}=p_{ij}+h_{ij}\text{ \ \ ,}  \tag{2.5}
\end{equation}%
together with the additional assumption that the two subspaces, on which the
projective tensor $p_{ij}$ and the tensor $h_{ij}$ are defined, are
orthogonal. This means that 
\begin{equation}
p_{ij}h^{jk}=0\text{ \ \ \ .}  \tag{2.6}
\end{equation}%
As a consequence 
\begin{equation}
p_{ij}p^{jk}=\delta _{i}^{k}-h_{ij}h^{jk}\neq \delta _{i}^{k}\text{ \ \ ,} 
\tag{2.7 }
\end{equation}%
meaning that the contravariant projective metric components $p^{jk}$ (in the
orthogonal subspace to the tensor $h_{ij}$) are no longer inverse to the
covariant ones $p_{ij}$.

An example of gravitational theories with more than one connection are the
so called \textit{theories with affine connections and metrics} [13], in
which there is one connection $\Gamma _{\alpha \beta }^{\gamma }$ for the
case of a parallel transport of covariant basic vectors $\nabla _{e_{\beta
}}e_{\alpha }=\Gamma _{\alpha \beta }^{\gamma }$ $e_{\gamma }$ and a \textit{%
separate} connection $P_{\alpha \beta }^{\gamma }$ for the contravariant
basic vector $e^{\gamma }$, the defining equation for which is $\nabla
_{e_{\beta }}e^{\alpha }=P_{\gamma \beta }^{\alpha }$ $e^{\gamma }$. In
these theories, the contravariant vector and tensor fields are assumed to be 
\textit{not the inverse ones} to the covariant vector and tensor fields.
This implies that 
\begin{equation}
e_{\alpha }e^{\beta }\equiv f_{\alpha }^{\beta }(x)\neq \delta _{\alpha
}^{\beta }  \tag{2.8}
\end{equation}%
and consequently, a distinction is made between \textit{covariant} and 
\textit{contravariant} metric tensors (and vectors too). Clearly, in the
above given case (2.7) of projective gravity, this theory should be
considered as a GTCCMC. In the same spirit, since the well - known Arnowitt
- Deser - Misner (ADM)\ (3+1) decomposition of spacetime [43, 44] is built
upon the projective transformation (2.5), it might be thought that it should
also be considered as such a theory. But in fact, the ADM (3+1) formalism
definitely is not an example for this \emph{due to the special
identification \ of the vector field's components [43, 44] with certain
components of the projective tensor, as a consequence of which all the
(spacelike defined) contravariant projective tensor components }$p^{\alpha
\beta }$\emph{\ (}$\alpha ,\beta ,\gamma =0,1,2,3$\emph{\ ; }$\ i,j=1,2,3$%
\emph{) turn out to be the inverse ones to the covariant projective
components }$p_{\alpha \gamma }\emph{.}$\emph{\ }In the case of the ADM
(3+1) decomposition, such an identification is indeed possible and
justified, since the gravitational field posseses coordinate invariance,
allowing to disentangle the dynamical degrees of freedom from the gauge
ones. But in the case when the tensors $h_{ij}$ are related with some moving
matter (with a prescribed motion) and an observer, "attached" to this matter
"measures" all the gravitational phenomena in his reference system by means
of the projective metric $p_{ij}$, this will be no longer possible. Then the
relation (2.7) will hold, and the resulting theory will be a\textit{\ GTCCMC.%
} Naturally, if the tensor $h_{ij}$ in (2.5) and (2.7) is taken in the form $%
h_{ij}=\frac{1}{e}u_{i}u^{j}$ and if the vector field $u$ (tangent at each
point of the trajectory of the moving matter) is assumed to be
non-normalized (i.e. $e(x)=u_{i}u^{i}\neq 1$), then one would have to work 
\textit{not within} the standard relativistic hydrodynamics theory (where $%
p_{ij}=g_{ij}-u_{i}u_{j}$ and $p_{ij}p^{jk}=\delta _{i}^{k}-u_{i}u^{k}$),
but within the formalism of GTCCMC (where $p_{ij}p^{jk}=f_{i}^{k}=\delta
_{i}^{k}-\frac{1}{e}u_{i}u^{k}\neq \delta _{i}^{k}$).\ One may wonder why
this should be so, since the last two formulaes for $p_{ij}p^{jk}$ for the
both cases look very much alike, with the exception of the "normalization"
function $\frac{1}{e}$ in the second formulae. But it shall become clear
that in the \textit{first case} the right-hand side has a \textit{tensor
transformation property}, while in the \textit{second case} due to the
function $\frac{1}{e}$ there would be no such property. And this shall turn
out to be crucial.

In order to understand this also from another point of view, let us perform
a covariant differentiation of both sides of the relation (2.8). Then one
can obtain that the two connections are related in the following way [13] 
\begin{equation}
f_{j,k}^{i}=\Gamma _{jk}^{l}\text{ }f_{l}^{i}+P_{lk}^{i}f_{j}^{l}\text{ \ \
\ \ \ ;\ \ \ \ \ \ (}f_{j,k}^{i}=\partial _{k}f_{j}^{i}\text{) \ \ \ \ . } 
\tag{2.13}
\end{equation}

Note also the following important moment - $f_{\alpha }^{\beta }(x)$ are
considered to be the components of a \textit{function. }Otherwise, if they
are considered to be a (mixed) tensor quantity, the covariant
differentiation of the mixed tensor $f_{\alpha }^{\beta }(x)$ in the
right-hand side of $e_{\alpha }e^{\beta }\equiv f_{\alpha }^{\beta }(x)$
would give exactly the same quantity as the one in the left-hand side for 
\textit{every choice} of the two connections $\Gamma _{jk}^{l}$ and $%
P_{lk}^{i}$, including also for the standard case (Einsteinian gravity) $%
P_{lk}^{i}=-\Gamma _{jk}^{l}$. This would mean that from a mathematical
point of view there would be no justification for the introduction of the
second covariant connection $P_{lk}^{i}$ . However, since  $f_{\alpha
}^{\beta }(x)$ are related with the description of some moving matter in the
Universe,  a tensor transformation law should not be prescribed to them. So
they should remain components of a function and consequently, the
introduction of the second connection $P_{lk}^{i}$ is inevitable.

In order  to understand further why and in \textit{what cases the
distinction between covariant and contravariant metric components will lead
to an inevitable introduction of two different connections }$\Gamma _{ij}^{k}
$\textit{\ and }$P_{ij}^{k}$, let us prove the following statement:

\begin{proposition}
If $e_{1},e_{2},...,e_{n}$ is a basis of covariant vector fields and $%
f_{i}^{\alpha }$ are the components of a given function or a constant, then
a basis of contravariant basic fields $\widetilde{e}^{\alpha _{1}},%
\widetilde{e}^{\alpha _{2}},...,\widetilde{e}^{\alpha _{n}}$ can be found so
that for each $i$ and $\alpha _{j}$ one has\ \ $e_{i}\widetilde{e}^{\alpha
_{j}}=f_{i}^{\alpha }$.
\end{proposition}

\bigskip This proposition is in fact is a generalization of the well-known
theorem from differential geometry that if a basis of (covariant) vector
fields is given, then a dual basis of (contravariant) vector fields can be
found, so that the contravariant vector fields are the \textit{inverse ones
to the covariant} ones, i.e. $e_{i}\widetilde{e}^{\alpha _{j}}=\delta
_{i}^{\alpha }$.

The proof is very simple, but essentially based on the relation (2.13). If
the covariant basic vector fields are given, then the contravariant
connection components $\Gamma _{ij}^{k}$ will be known too. Since $%
f_{j,k}^{i}$ are derivatives of a function, one may take the expression
(2.13) \ \ \ \ \ \ \ \ $f_{j,k}^{i}=\Gamma _{jk}^{l}$ $%
f_{l}^{i}+P_{lk}^{i}f_{j}^{l}$ , which for the moment shall be treated as a
system of $n.[\frac{n(n+1)}{2}]$ \textit{linear algebraic equations with
respect to the (unknown) connection components }$P_{lk}^{i}$. A solution of
this system can be found for the connection components $P_{lk}^{i}$. Then
the condition for the parallel transport of the contravariant basic vector
fields $\nabla _{e_{\beta }}\widetilde{e}^{\alpha }=P_{\gamma \beta
}^{\alpha }$ $\widetilde{e}^{\gamma }$ can be written as $\partial _{\beta }%
\widetilde{e}^{\alpha }=P_{\gamma \beta }^{\alpha }$ $\widetilde{e}^{\gamma }
$ and considered as a system of $n$\textit{\ ordinary differential equations
with respect to the components }$\widetilde{e}^{\alpha }$. From the solution
of this system, $\widetilde{e}^{\alpha }$ can be found. So the new basis $%
\widetilde{e}^{\alpha _{1}},\widetilde{e}^{\alpha _{2}},...,\widetilde{e}%
^{\alpha _{n}}$ will  be \textit{\ uniquely determined, up to the
integration constants, }contained in the solution of the system of
differential equations. 

After proving this proposition, the difference between standard relativistic
hydrodynamics and \textit{"modified" relativistic hydrodynamics with a
variable length} can be easily understood. In the first case, the right-hand
side in $p_{ij}p^{jk}=\delta _{i}^{k}-u_{i}u^{k}=f_{i}^{k}\neq \delta
_{i}^{k}$ transforms as a tensor, which is ensured also by normalization
property $u_{i}u^{i}=1$.Therefore (2.13) and the proposition will not hold,
so the contravariant basic vector fields are determined in the standard way $%
e_{i}e^{j}=\delta _{i}^{j}$ and more importantly, they \textit{cannot be
determined} in another way, in spite of the fact that  $f_{i}^{k}\neq \delta
_{i}^{k}$.

In the second case, the situation is just the opposite - the right-hand side
of $p_{ij}p^{jk}=\delta _{i}^{k}-\frac{1}{e}u_{i}u^{k}=f_{i}^{k}\neq \delta
_{i}^{k}$ transforms not as a tensor because of the "normalization" factor $%
\frac{1}{e}$, the proposition holds and thus the basic vector fields are
determined as $e_{i}\widetilde{e}^{j}=f_{i}^{j}$. \textit{Therefore, the
treatment of relativistic hydrodynamics with "variable length" should be
within the GTCCMC}.

In the present case, the introduced new connection (2.4) \textit{should not
be identified} with the connection $P_{\alpha \beta }^{\gamma }$, since the
connection $\widetilde{\Gamma }_{kl}^{s}\equiv \widetilde{g}^{is}\Gamma
_{i;kl}$ \ is \ introduced by means of modifying the contravariant tensor
and not on the base of any separately defined  parallel transport for the
contravariant basic vectors. Moreover, the connection $\widetilde{\Gamma }%
_{kl}^{s}$ turns out to be a linear combination of the Christoffell
connection components \ $\Gamma _{\alpha \beta }^{\gamma }$, and the
relation between them is not of the type (2.13). In such a way, there will
not be a contradiction with the case when the\textit{\ two connections }$%
\Gamma _{\alpha \beta }^{\gamma }$\textit{\ and }$\widetilde{\Gamma }%
_{kl}^{s}$ \textit{are not defined as separate ones}, since later on, in
deriving the cubic algebraic equation in the general case and for the case $%
\widetilde{g}^{jk}=dX^{j}dX^{k}$ also, it would be supposed that $\widetilde{%
g}^{is}$ is a tensor. This would mean (from $\widetilde{g}^{is}g_{im}\equiv
f_{m}^{s}(\mathbf{x})$) that $f_{m}^{s}(\mathbf{x})$ will also be a (mixed)
tensor quantity, and therefore the covariant differentiation of $e_{\alpha
}e^{\beta }\equiv f_{\alpha }^{\beta }(x)$ will not produce any new relation.

\section{\protect\bigskip\ BASIC \ ALGEBRAIC \ EQUATIONS \ IN \ GRAVITY \
THEORY. TENSOR\ \ LENGTH\ SCALE}

If one writes down the Ricci tensor in terms of the newly defined
contravariant tensor $\widetilde{g}^{ij}\equiv dX^{i}dX^{j}$, then the
following fourth - degree algebraic equation can be obtained 
\begin{equation*}
R_{ik}=dX^{l}\left[ g_{is,l}\frac{\partial (dX^{s})}{\partial x^{k}}-\frac{1%
}{2}pg_{ik,l}+\frac{1}{2}g_{il,s}\frac{\partial (dX^{s})}{\partial x^{k}}%
\right] +
\end{equation*}%
\begin{equation}
+\frac{1}{2}dX^{l}dX^{m}dX^{r}dX^{s}\left[
g_{m[k,t}g_{l]r,i}+g_{i[l,t}g_{mr,k]}+2g_{t[k,i}g_{mr,l]}\right] \text{ \ \
\ \ \ ,}  \tag{3.1}
\end{equation}%
where $p$ is the scalar quantity

\begin{equation}
p\equiv div(dX)\equiv \frac{\partial (dX^{l})}{\partial x^{l}}\text{,} 
\tag{3.2}
\end{equation}%
which \textquotedblright measures\textquotedblright\ the divergency of the
vector field $dX$. The algebraic variety of the equation consists of the
differentials $dX^{i\text{ }}$ and their derivatives $\frac{\partial (dX^{s})%
}{\partial x^{k}}$.

In the same spirit, one can investigate the problem whether the
gravitational Lagrangian in terms of the new contravariant tensor can be
equal to the standard representation of the gravitational Lagrangian. This
standard \textit{(first) representation} of the gravitational Lagrangian is
based on the standard Christoffell connection $\Gamma _{ij}^{k}$ (given by
formulae (2.1)), the Ricci tensor $R_{ik}$ (formulae (2.2)) and the \textit{%
other contravariant tensor} $\widetilde{g}^{ij}=dX^{i}dX^{j}$ [10] 
\begin{equation}
L_{1}=-\sqrt{-g}\widetilde{g}^{ik}R_{ik}=-\sqrt{-g}dX^{i}dX^{k}R_{ik}\text{
\ .}  \tag{3.3}
\end{equation}%
In the \textit{second representation, }the\textit{\ }Christoffell connection 
$\widetilde{\Gamma }_{ij}^{k}$ and the Ricci tensor $\widetilde{R}_{ik}$%
\textit{\ }are "tilda" quantities, meaning that the "tilda" Christoffell
connection is determined by formulae (2.4) with the new contravariant tensor 
$\widetilde{g}^{ij}=dX^{i}dX^{j}$ and the "tilda" Ricci tensor $\widetilde{R}%
_{ik}$ - by formulae (2.2), but with the "tilda" connection $\widetilde{%
\Gamma }_{ij}^{k}$ instead of the usual Christoffell connection $\Gamma
_{ij}^{k}$. Thus the expression for the \textit{second representation} of
the gravitational Lagrangian acquires the form 
\begin{equation}
L_{2}=-\sqrt{-g}\widetilde{g}^{il}\widetilde{R}_{il}=-\sqrt{-g}%
dX^{i}dX^{l}\{p\Gamma _{il}^{r}g_{kr}dX^{k}-\Gamma
_{ik}^{r}g_{lr}d^{2}X^{k}-\Gamma _{l(i}^{r}g_{k)r}d^{2}X^{k}\}\text{ .} 
\tag{3.4}
\end{equation}%
The condition for the \textit{equivalence of the two representations} $%
L_{1}=L_{2}$ gives a cubic algebraic equation with respect to the algebraic
variety of the first differential $dX^{i}$ and the second ones $d^{2}X^{i}$
[10] 
\begin{equation}
dX^{i}dX^{l}\left( p\Gamma _{il}^{r}g_{kr}dX^{k}-\Gamma
_{ik}^{r}g_{lr}d^{2}X^{k}-\Gamma _{l(i}^{r}g_{k)r}d^{2}X^{k}\right)
-dX^{i}dX^{l}R_{il}=0\text{ \ \ \ \ .}  \tag{3.5}
\end{equation}
In [22] also another cubic algebraic equation has been obtained, but after
the application of a variational approach.

\textbf{\ }Following the  approach in [10], the Einstein's equations in
vacuum for the general case were derived under the assumption that the
contravariant metric tensor components are the "tilda" ones: 
\begin{equation*}
0=\widetilde{R}_{ij}-\frac{1}{2}g_{ij}\widetilde{R}=\widetilde{R}_{ij}-\frac{%
1}{2}g_{ij}dX^{m}dX^{n}\widetilde{R}_{mn}=
\end{equation*}%
\begin{equation*}
=-\frac{1}{2}pg_{ij}\Gamma _{mn}^{r}g_{kr}dX^{k}dX^{m}dX^{n}+\frac{1}{2}%
g_{ij}(\Gamma _{km}^{r}g_{nr}+\Gamma
_{n(m}^{r}g_{k)r})d^{2}X^{k}dX^{m}dX^{n}+
\end{equation*}%
\begin{equation}
+p\Gamma _{ij}^{r}g_{kr}dX^{k}-(\Gamma _{ik}^{r}g_{jr}+\Gamma
_{j(i}^{r}g_{k)r})d^{2}X^{k}\text{ \ \ \ .}  \tag{3.6 }
\end{equation}

\bigskip This equation represents again a system of cubic equations. In
addition, if the  differentials $dX^{i}$\ and $d^{2}X^{i}$\ are known, but
not the covariant tensor $g_{ij}$, the same equation can be considered also
as a \ cubic algebraic equation with respect to the algebraic variety of the
metric tensor components $g_{ij}$\ and their first derivatives $g_{ij,k}$.

It might be thought that the definite choice of the contravariant tensor in
the form of the factorized product $\widetilde{g}^{ij}=dX^{i}dX^{j}$ is a
serious restriction, in view of the fact that the second derivatives of the
covariant tensor components $g_{ij,kl}$ are not present in the equation.
This is indeed so, because the algebraic structure of the equation is
simpler to deal with in comparison with the general case, and so it is
easier to implement the algorithm for parametrization, developed in [10].
But there is one argument in favour of this choice (although the case for an
arbitrary contravariant tensor is no doubt more important) - since the
metric can be expressed as $ds^{2}=l(x)=g_{ij}dX^{i}dX^{j}$ (consequently $%
dX^{i}dX^{j}=l(x)g^{ij}$), the obtained cubic algebraic equations (3.5) and
(3.6) can be considered in regard also to the length function $l(x)$. Since
for Einsteinian gravity $g_{ij}g^{jk}=\delta _{i}^{k}$ (i.e. $g^{jk}=%
\widetilde{g}^{jk}=dX^{j}dX^{k}$), then for this case the length function is
"postulated" to be $l=1$. But the length function can also be obtained as a
solution of the cubic equation, and thus in more general theories of gravity
solutions with $l\neq 1$ may exit. In fact, for a general contravariant
tensor $\widetilde{g}^{ij}\neq dX^{i}dX^{j}$, one would have $\widetilde{g}%
^{ij}=l_{k}^{i}g^{kj}$, where $l_{k}^{i}$ will be proposed to be called a 
\textit{"tensor length scale", }and the previously defined length function $%
l(x)$ is a partial case of the tensor length scale for $l_{j}^{i}=l\delta
_{j}^{i}$. The \textit{physical meaning} of the notion of tensor length
scale is simple - in the different directions (i.e. for different $i$ and $j$%
) the length scale is \ different. In particular, some motivation for this
comes from Witten's paper [45], where in discussing some aspects of weakly
coupled heterotic string theory (when there is just one string couplings )
and the obtained too large bound on the Newton's constant it was remarked
that \textit{\textquotedblright the problem might be ameliorated by
considering an anisotropic Calabi - Yau with a scale }$\sqrt{\alpha
^{^{\prime }}}$\textit{\ in }$d$\textit{\ directions and }$\frac{1}{M_{GUT}}$%
\textit{\ in }$(6-d)$\textit{\ directions\textquotedblright }. Thus it may
be proposed to realize this if one takes 
\begin{equation}
l_{i}^{k}=g_{ij}dX^{j}dX^{k}=l_{1}\delta _{i}^{k}\text{ \ for \ }%
i,j,k=1,....,d\text{ \ \ \ \ ,}  \tag{3.7}
\end{equation}%
\begin{equation}
l_{a}^{b}=g_{ac}dX^{c}dX^{b}=l_{2}\delta _{a}^{b}\text{ \ for \ }%
a,b,c=d+1,....,6\text{\ \ \ \ .}  \tag{3.8}
\end{equation}%
Note also the justification for the name \textit{"tensor length scale"} - if
\ $l_{k}^{i}$ is a tensor quantity, so will be the "modified" contravariant
tensor $\widetilde{g}^{ij}=l_{k}^{i}g^{kj}$, and consequently in accord with
section 2 there will be no need for the introduction of a new covariant
connection $P_{ij}^{k}$. And this is indeed the case, because the relation
between the two connections $\Gamma _{ij}^{k}$ and $\widetilde{\Gamma }%
_{ij}^{k}$ is given by formulae (2.4) $\widetilde{\Gamma }_{kl}^{s}:=%
\widetilde{g}^{is}g_{im}\Gamma _{kl}^{m}$. In other words, these two
connections are not considered to be "separately introduced" and so they do
not depend on one another by means of the equality (2.13). 

The purpose of the present paper further will be:\ \textit{how can one
extend the proposed in [10] approach for the "modified" contravariant metric
components (as }$\widetilde{g}^{ij}=dX^{i}dX^{j}$\textit{) to the case of a
generally defined contravariant tensor }$\widetilde{g}^{ij}\neq dX^{i}dX^{j}$%
\textit{? }

\section{ INTERSECTING  ALGEBRAIC  VARIETIES AND STANDARD (EINSTEINIAN)
GRAVITY THEORY}

A more general theory with the definition of the contravariant tensor as $%
\widetilde{g}^{ij}\equiv dX^{i}dX^{j}$ should \ contain in itself the
standard gravitational theory with $g_{ij}g^{jk}=\delta _{i}^{k}$. From a
mathematical point of view, this should be performed by considering the
intersection [19, 20, 21] of the cubic algebraic equations (3.6)\ \ with the
system of \ $n^{2}$ quadratic algebraic equations for the algebraic variety
of the $n$ variables 
\begin{equation}
g_{ij}dX^{j}dX^{k}=\delta _{i}^{k}\text{ \ \ .}  \tag{4.1}
\end{equation}%
In its general form $g_{ij}\widetilde{g}^{jk}=\delta _{i}^{k}$ with an
arbitrary contravariant tensor $\widetilde{g}^{jk}$, this system \ can also
be considered together with the Einstein's "algebraic" system of equations,
which in the next section shall be derived for a \textit{generally defined
contravariant tensor.} From an algebro - geometric point of view, this is
the problem about the intersection of the Einstein's algebraic equations
with the system of $n^{2}$ \ (linear) hypersurfaces for the $\left[ \left( 
\begin{array}{c}
n \\ 
2%
\end{array}%
\right) +n\right] $\ contravariant variables, if the covariant tensor
components are given. Since the derived Einstein's algebraic equations are
again cubic ones with respect to the contravariant metric components, this
is an analogue to the well - known problem in algebraic geometry about the
intersection of a (two-dimensional)\ cubic curve with a straight line.
However, in the present case the straight line and the cubic curve are 
\textit{multi - dimensional ones}, which is a substantial difference from
the standard case in algebraic geometry.

The standardly known solutions of the Einstein's equations can be obtained
as an intersection variety of the Einstein's algebraic equations with the
system $g_{ij}\widetilde{g}^{jk}=\delta _{i}^{k}$ . However, the strict
mathematical proof that such an intersection will give the known solutions
is still lacking.

\section{ ALGEBRAIC \ EQUATIONS \ FOR \ A \ GENERAL \ CONTRAVARIANT \ METRIC
\ TENSOR}

Let us write down the algebraic equations for all admissable
parametrizations of the gravitational Lagrangian for the generally defined
contravariant tensor $\widetilde{g}^{ij}$, following \textit{the same
prescription} as in section 3, where the equality of the two representations
of the gravitational Lagrangian has been supposed: 
\begin{equation*}
\widetilde{g}^{i[k}\widetilde{g}_{,l}^{l]s}\Gamma _{ik}^{r}g_{rs}+\widetilde{%
g}^{i[k}\widetilde{g}^{l]s}\left( \Gamma _{ik}^{r}g_{rs}\right) _{,l}+
\end{equation*}%
\begin{equation}
+\widetilde{g}^{ik}\widetilde{g}^{ls}\widetilde{g}^{mr}g_{pr}g_{qs}\left(
\Gamma _{ik}^{q}\Gamma _{lm}^{p}-\Gamma _{il}^{p}\Gamma _{km}^{q}\right) -R=0%
\text{ \ \ \ \ .}  \tag{5.1}
\end{equation}%
This equation is again a \textit{cubic algebraic equation} with \ respect to
the algebraic variety of the variables $\widetilde{g}^{ij}$ and $\widetilde{g%
}_{,k}^{ij}$, and the number of variables in the present case is much
greater than in the previous case for the contravariant tensor $\widetilde{g}%
^{ij}\equiv dX^{i}dX^{j}$ . At the same time, this equation is a \textit{%
fourth - degree} algebraic equation with respect to the covariant metric
tensor $g_{ij}$ and its first and second partial derivatives. With respect
to the algebraic variety of all the variables $\widetilde{g}^{ij}$, $%
\widetilde{g}_{,k}^{ij}$, $g_{ij}$, $g_{ij,k}$, $g_{ij,kl}$, the above
algebraic equation is of \textit{seventh order} and with coefficient
functions, due to the presence of the terms with the affine connection $%
\Gamma _{ik}^{q}$ and its derivatives, which contain the contravariant
tensor $g^{ij}$ and $g_{,k}^{ij}$.

Similarly, the Einstein's equations can be written as a system of \ \textit{%
third - degree  algebraic equations} with respect to the (generally chosen)
contravariant variables and their derivatives 
\begin{equation*}
0=\widetilde{R}_{ij}-\frac{1}{2}g_{ij}\widetilde{R}=
\end{equation*}%
\begin{equation*}
=\widetilde{g}^{lr}(\Gamma _{r;i[j}),_{l]}+\widetilde{g}_{,[l}^{lr}\Gamma
_{r;ij]}+\widetilde{g}^{lr}\widetilde{g}^{ms}(\Gamma _{r;ij}\Gamma
_{s;lm}-\Gamma _{s;il}\Gamma _{r;km})-
\end{equation*}%
\begin{equation*}
-\frac{1}{2}g_{ij}\widetilde{g}^{m[k}\widetilde{g}_{,l}^{l]s}\Gamma
_{mk}^{r}g_{rs}-\frac{1}{2}g_{ij}\widetilde{g}^{m[k}\widetilde{g}%
^{l]s}\left( \Gamma _{mk}^{r}g_{rs}\right) _{,l}-
\end{equation*}%
\begin{equation}
-\frac{1}{2}g_{ij}\widetilde{g}^{nk}\widetilde{g}^{ls}\widetilde{g}%
^{mr}g_{pr}g_{qs}\left( \Gamma _{nk}^{q}\Gamma _{lm}^{p}-\Gamma
_{nl}^{p}\Gamma _{km}^{q}\right) \text{ \ .}  \tag{5.2}
\end{equation}%
Interestingly, the same system of equations can be considered as a system of 
\textit{fifth - degree} equations with respect to the covariant variables
(which is the difference from the previous case). The mathematical treatment
of fifth - degree equations is known since the time of Felix Klein's famous
monograph [24], published in 1884. A way for resolution of such equations on
the base of earlier developed approaches by means of reducing the fifth -
degree equations to the so called modular equation has been presented in the
more recent \ monograph of \ Prasolov and Solov'yev [9]. Some other methods
for solution of third-, fifth- and higher- order \ algebraic equations have
been given in [25, 26]. A complete description of elliptic, theta and
modular functions has been given in the old monographs [27, 28]. Also,
solutions of $n-$ th degree algebraic equations in theta - constants [29]
and in special functions [30] are interesting in view of the not yet proven
hypothesis in the paper by Kraniotis and Whitehouse [8] that \textit{%
\textquotedblright all nonlinear solutions of general relativity are
expresed in terms of theta - functions, associated with Riemann -
surfaces\textquotedblright }. Some other monographs, related to elliptic
functions and elliptic curves are [31-42]. 

Two other important problems can be pointed out:

1. One can find solutions of the system of Einstein's equations not as
solutions of a system of nonlinear differential equations, but as elements
of an algebraic variety, satisfying the Einstein's algebraic equations. The
important new moment is that this gives an opportunity to find solutions of
\ the Einstein's equations both for the components of the covariant metric
tensor $g_{ij}$\ and for the contravariant ones $\widetilde{g}^{jk}$. This
means that solutions may exist for which $g_{ij}\widetilde{g}^{jk}\neq
\delta _{i}^{k}$. In other words, a classification of the solutions of the
Einstein's equations can be performed in an entirely new and nontrivial
manner - under a given contravariant tensor, the covariant tensor and its
derivatives have to be found from the algebraic equation, or under a given
covariant tensor, the contravariant tensor and its derivatives can be found.

2. The condition for the zero - covariant derivative of the covariant metric
tensor $\nabla _{k}g_{ij}=0$ and of the contravariant metric tensor $\nabla
_{k}\widetilde{g}^{ij}=0$ can be written in the form of the following cubic
algebraic equations\textbf{\ }with respect to the variables $g_{ij}$, $%
g_{ij,k}$ and $\widetilde{g}^{ls}$ \textbf{:} 
\begin{equation}
\nabla _{k}g_{ij}\equiv g_{ij,k}-\widetilde{\Gamma }%
_{k(i}^{l}g_{j)l}=g_{ij,k}-\widetilde{g}^{ls}\Gamma _{s;k(i}g_{j)l}=0 
\tag{5.3}
\end{equation}%
and 
\begin{equation}
0=\nabla _{k}\widetilde{g}^{ij}=\widetilde{g}_{,k}^{ij}+\widetilde{g}^{r(i}%
\widetilde{g}^{j)s}\Gamma _{r;sk}\text{ \ \ \ \ .}  \tag{5.4}
\end{equation}%
The first equation (5.3) is linear with respect to $\widetilde{g}^{ls}$ and
quadratic with respect to $g_{ij}$, $g_{ij,k}$, while the second equation
(5.3) is linear with respect to $g_{ij}$, $g_{ij,k}$ and quadratic with
respect to $\widetilde{g}^{ls}$.

\section{\protect\bigskip\ EMBEDDED \ SEQUENCE \ OF \ ALGEBRAIC \ EQUATIONS
\ AND \ FINDING \ THE \ SOLUTIONS \ OF \ \ THE \ CUBIC \ ALGEBRAIC \ EQUATION%
}

The purpose of the present subsection will be to describe the method for
finding the solution (i. e . the algebraic variety of the differentials $%
dX^{i}$) of the cubic algebraic equation (3.5) (in the limit $d^{2}X^{k}=0$%
). The method has been proposed first in [10] but here it will be developed
further and applied with respect to a \textit{sequence of algebraic equations%
} with algebraic varieties, which are embedded into the initial one. This
means that if at first the algorithm is applied with respect to the
three-dimensional cubic algebraic equation (3.5) and a solution for $dX^{3}$
(depending on the Weierstrass function and its derivative) is found, then
the same algorithm will be applied with respect to the two-dimensional cubic
algebraic equation with variables $dX^{1}$\ and $dX^{2}$, and finally to the
one-dimensional cubic algebraic equation of the variable $dX^{1}$\ only.

The basic and very simple idea about parametrization of a cubic algebraic
equation with the Weierstrass function [9, 11,12] can be presented as
follows: Let us define the lattice $\Lambda =\{m\omega _{1}+n\omega _{2}\mid
m,n\in Z;$ $\omega _{1},\omega _{2}\in C,Im\frac{\omega _{1}}{\omega _{2}}%
>0\}$ and the mapping $f:$ $C/\Lambda \rightarrow CP^{2}$, which maps the
factorized (along the points of the lattice $\Lambda $) part of the points
on the complex plane into the two dimensional complex projective space $%
CP^{2}$. This means that each point $z$ on the complex plane is mapped onto
the point $(x,y)=(\rho (z),\rho ^{^{\prime }}(z))$, where $x$ and $y$ belong
to the affine curve 
\begin{equation}
y^{2}=4x^{3}-g_{2}x-g_{3}\text{ \ \ \ \ \ \ .\ }  \tag{6.1}
\end{equation}%
In other words, the functions $x=\rho (z)$ and $y=\rho ^{^{\prime }}(z)$,
where $\rho (z)$ denotes the \textit{Weierstrass elliptic function} (1.3),
are \textit{uniformization functions} for the cubic curve. It can be proved
[9] that the only cubic algebraic curve with number coefficients, which is
parametrized by the uniformization functions $x=\rho (z)$ and $y=\rho
^{^{\prime }}(z)$ is the affine curve (6.1).  

In the case of the cubic equation of reparametrization invariance (3.5), the
aim will be again to bring the equation to the form (6.1) and afterwards to
make equal each of the coefficient functions to the (numerical) coefficients
in (6.1).

In order to provide a more clear description of the developed method, let us
divide it into several steps.

\textbf{Step 1}. The initial cubic algebraic equation (3.5) is presented as
a cubic equation with respect to the variable $dX^{3}$ only 
\begin{equation}
A_{3}(dX^{3})^{3}+B_{3}(dX^{3})^{2}+C_{3}(dX^{3})+G^{(2)}(dX^{2},dX^{1},g_{ij},\Gamma _{ij}^{k},R_{ik})\equiv 0%
\text{ \ \ \ \ ,}  \tag{6.2}
\end{equation}%
where naturally the coefficient functions $A_{3}$, $B_{3}$ , $C_{3}$ and $%
G^{(2)}$ depend on the variables $dX^{1}$ and $dX^{2}$ of the algebraic
subvariety and on the metric tensor $g_{ij}$, the Christoffel connection $%
\Gamma _{ij}^{k}$ and the Ricci tensor $R_{ij}$: 
\begin{equation}
A_{3}\equiv 2p\Gamma _{33}^{r}g_{3r}\text{ \ \ ; \ \ \ \ \ \ \ \ \ \ \ }%
B_{3}\equiv 6p\Gamma _{\alpha 3}^{r}g_{3r}dX^{\alpha }-R_{33}\text{\ \ \ \
,\ }  \tag{6.3}
\end{equation}%
\begin{equation}
C_{3}\equiv -2R_{\alpha 3}dX^{\alpha }+2p(\Gamma _{\alpha \beta
}^{r}g_{3r}+2\Gamma _{3\beta }^{r}g_{\alpha r})dX^{\alpha }dX^{\beta }\text{
\ \ \ .}  \tag{6.4}
\end{equation}%
The Greek indices $\alpha ,\beta $ take values $\alpha ,\beta =1,2$ while
the indice $r$ takes values $r=1,2,3$.

\textbf{Step 2}. A linear-fractional transformation 
\begin{equation}
dX^{3}=\frac{a_{3}(z)\widetilde{dX}^{3}+b_{3}(z)}{c_{3}(z)\widetilde{dX}%
^{3}+d_{3}(z)}  \tag{6.5 }
\end{equation}%
is performed with the purpose of setting up to zero the coefficient
functions in front of the highest (third) degree of $\ \widetilde{dX}^{3}$.
This will be achieved if $G^{(2)}(dX^{2},dX^{1},g_{ij},\Gamma
_{ij}^{k},R_{ik})=-\frac{a_{3}Q}{c_{3}^{3}}$, where 
\begin{equation}
Q\equiv A_{3}a_{3}^{2}+C_{3}c_{3}^{2}+B_{3}a_{3}c_{3}+2c_{3}d_{3}C_{3}\text{
\ \ \ \ \ \ . }  \tag{6.6 }
\end{equation}%
This \ gives a cubic algebraic equation with respect to the two-dimensional
algebraic variety of the variables $dX^{1}$ and $dX^{2}$: 
\begin{equation}
p\Gamma _{\gamma (\alpha }^{r}g_{\beta )r}dX^{\gamma }dX^{\alpha }dX^{\beta
}+K_{\alpha \beta }^{(1)}dX^{\alpha }dX^{\beta }+K_{\alpha }^{(2)}dX^{\alpha
}+2p\left( \frac{a_{3}}{c_{3}}\right) ^{3}\Gamma _{33}^{r}g_{3r}=0\text{ \ \
\ }  \tag{6.7 }
\end{equation}%
and $K_{\alpha \beta }^{(1)}$ and $K_{\alpha }^{(2)}$ are the corresponding
quantities [10] 
\begin{equation}
K_{\alpha \beta }^{(1)}\equiv -R_{\alpha \beta }+2p\frac{a_{3}}{c_{3}}(1+2%
\frac{d_{3}}{c_{3}})(2\Gamma _{\alpha \beta }^{r}g_{3r}+\Gamma _{3\alpha
}^{r}g_{\beta r})\text{ \ \ \ }  \tag{6.8 }
\end{equation}%
and 
\begin{equation}
K_{\alpha }^{(2)}\equiv 2\frac{a_{3}}{c_{3}}\left[ 3p\frac{a_{3}}{c_{3}}%
\Gamma _{\alpha 3}^{r}g_{3r}-(1+2\frac{d_{3}}{c_{3}})R_{\alpha 3}\right] 
\text{ \ \ .}  \tag{6.9}
\end{equation}%
Note that since the linear fractional transformation (with another
coefficient functions) will again be applied with respect to another cubic
equations, everywhere in (6.5 - 6.8) the coefficient functions $a_{3}(z)$, $%
b_{3}(z)$, $c_{3}(z)$ and $d_{3}(z)$ bear the indice \textquotedblright $3$%
\textquotedblright ,\ to distinguish them from the indices in the other
linear-fractional tranformations, which are to be applied. In terms of the
new variable $n_{3}=\widetilde{dX}^{3}$ the original cubic equation (3.5)\
acquires the form [10] 
\begin{equation}
\widetilde{n}^{2}=\overline{P}_{1}(\widetilde{n})\text{ }m^{3}+\overline{P}%
_{2}(\widetilde{n})\text{ }m^{2}+\overline{P}_{3}(\widetilde{n})\text{ }m+%
\overline{P}_{4}(\widetilde{n})\text{ ,}  \tag{6.10 }
\end{equation}%
where $\overline{P}_{1}(\widetilde{n})$ $,\overline{P}_{2}(\widetilde{n}),$ $%
\overline{P}_{3}(\widetilde{n})$ and $\overline{P}_{4}(\widetilde{n})$ are
complicated functions of the ratios $\frac{c_{3}}{d_{3}}$, $\frac{b_{3}}{%
d_{3}}$ and $A_{3},B_{3},C_{3}$ (but not of the ratio $\frac{a_{3}}{d_{3}}$,
which is very important). The variable $m$ denotes the ratio $\frac{a_{3}}{%
c_{3}}$ and the variable $\widetilde{n}$ is related to the variable $n$
through the expresssion 
\begin{equation}
\widetilde{n}=\sqrt{k_{3}}\sqrt{C_{3}}\left[ n+L_{1}^{(3)}\frac{B_{3}}{C_{3}}%
+L_{2}^{(3)}\right] \text{ \ \ \ \ \ ,}  \tag{6.11}
\end{equation}%
where 
\begin{equation}
k_{3}\equiv \frac{b_{3}}{d_{3}}\frac{c_{3}}{d_{3}}(\frac{c_{3}}{d_{3}}+2)%
\text{ \ \ \ \ \ ,}  \tag{6.12}
\end{equation}%
\begin{equation}
L_{1}^{(3)}\equiv \frac{1}{2}\frac{\frac{b_{3}}{d_{3}}}{\frac{c_{3}}{d_{3}}+2%
}\text{ \ \ ; \ \ \ }L_{2}^{(3)}\equiv \frac{1}{\frac{c_{3}}{d_{3}}+2}\text{
\ \ .}  \tag{6.13}
\end{equation}

The subscript \textquotedblright $3$\textquotedblright\ in $L_{1}^{(3)}$ and 
$L_{2}^{(3)}$ means that the corresponding ratios in the right-hand side 
also have the same subscript. Setting up the coefficient functions $%
\overline{P}_{1}(\widetilde{n})$ $,\overline{P}_{2}(\widetilde{n}),$ $%
\overline{P}_{3}(\widetilde{n})$ equal to the number coefficients $%
4,0,-g_{2},-g_{3}$ respectively, one can now parametrize the resulting
equation 
\begin{equation}
\widetilde{n}^{2}=4m^{3}-g_{2}m-g_{3}\text{ }  \tag{6.14 }
\end{equation}%
according to the standard prescription 
\begin{equation}
\widetilde{n}=\rho ^{^{\prime }}(z)=\frac{d\rho }{dz}\text{ \ \ \ \ \ \ \ \
\ \ \ ;\ \ \ \ \ \ \ \ \ \ \ \ \ \ }\frac{a_{3}}{c_{3}}\equiv \text{\ }%
m=\rho (z)\text{\ \ \ \ .}  \tag{6.15 }
\end{equation}%
Taking this into account, representing the linear-fractional transformation
(6.5) as (dividing by $\ c_{3}$) 
\begin{equation}
dX^{3}=\frac{\frac{a_{3}}{c_{3}}\widetilde{dX}^{3}+\frac{b_{3}}{c_{3}}}{%
\widetilde{dX}^{3}+\frac{d_{3}}{c_{3}}}  \tag{6.16 }
\end{equation}%
and combining expressions (6.11) for $\widetilde{n}$ and (6.16), one can
obtain the final formulae for $dX^{3}$ as a solution of the cubic algebraic
equation 
\begin{equation}
dX^{3}=\frac{\frac{b_{3}}{c_{3}}+\frac{\rho (z)\rho ^{^{\prime }}(z)}{\sqrt{%
k_{3}}\sqrt{C_{3}}}-L_{1}^{(3)}\frac{B_{3}}{C_{3}}\rho (z)-L_{2}^{(3)}\rho
(z)}{\frac{d_{3}}{c_{3}}+\frac{\rho ^{^{\prime }}(z)}{\sqrt{k_{3}}\sqrt{C_{3}%
}}-L_{1}^{(3)}\frac{B_{3}}{C_{3}}-L_{2}^{(3)}}\text{ \ \ \ \ \ .}  \tag{6.17}
\end{equation}%
In order to be more precise, it should be mentioned that the identification
of the functions $\overline{P}_{1}(\widetilde{n})$ $,\overline{P}_{2}(%
\widetilde{n}),$ $\overline{P}_{3}(\widetilde{n})$ with the number
coefficients gives some additional equations [10], which in principle have
to be taken into account in the solution for $dX^{3}$. This has been
investigated to a certain extent in [10], and will be continued to be
investigated. Here in this paper the main objective will be to show the
dependence of the solutions on the Weierstrass function and its derivative.
Since only the ratios $\frac{b}{d}$ and $\frac{c}{d}$ enter these additional
relations, and not $\frac{a}{c}$ (which is related to the Weierstrass
function), they do not affect the solution with respect to $\rho (z)$ and $%
\rho ^{^{\prime }}(z)$.

Since $B_{3}$ and $C_{3}$ depend on $dX^{1}$ and $dX^{2}$, the solution
(6.17)\ for $dX^{3}$ shall be called \textit{the embedding solution} for $%
dX^{1}$ and $dX^{2}$.

\textbf{Step 3.} Let us now consider the two-dimensional cubic equation
(6.6). Following the same approach and finding the \textquotedblright
reduced\textquotedblright\ cubic algebraic equation for $dX^{1}$ only, it
shall be proved that the solution for $dX^{2}$ is the embedding solution for 
$dX^{1}$.

For the purpose, let us again write down eq. (6.6)\ in the form (6.2),
singling out the variable $dX^{2}$: 
\begin{equation}
A_{2}(dX^{2})^{3}+B_{2}(dX^{2})^{2}+C_{2}(dX^{2})+G^{(1)}(dX^{1},g_{ij},%
\Gamma _{ij}^{k},R_{ik})\equiv 0\text{ \ \ \ \ ,}  \tag{6.18}
\end{equation}%
where the coefficient functions $A_{2},B_{2},C_{2}$ and $G^{(1)}$ are the
following: 
\begin{equation}
A_{2}\equiv 2p\Gamma _{22}^{r}g_{2r}\text{ \ \ ; \ \ \ \ \ \ \ \ \ \ \ }%
B_{2}\equiv K_{22}^{(1)}+2p[2\Gamma _{12}^{r}g_{2r}+\Gamma
_{22}^{r}g_{1r}]dX^{1}\text{\ \ \ \ ,\ }  \tag{6.19 }
\end{equation}%
\begin{equation}
C_{2}\equiv 2p[\Gamma _{11}^{r}g_{2r}+2\Gamma
_{12}^{r}g_{1r})(dX^{1})^{2}+(K_{12}^{(1)}+K_{21}^{(1)})dX^{1}+K_{2}^{(2)}%
\text{ \ \ \ ,}  \tag{6.20 }
\end{equation}%
\begin{equation}
G^{1}\equiv 2p\Gamma
_{11}^{r}g_{1r}(dX^{1})^{3}+K_{11}^{(1)}(dX^{1})^{2}+K_{1}^{(2)}dX^{1}+2p%
\rho ^{3}(z)\Gamma _{33}^{r}g_{3r}\text{ \ \ .}  \tag{6.21 }
\end{equation}%
Note that the starting equation (6.7) has the same structure of the first
terms, if one makes the formal substitution $-R_{\alpha \beta }\rightarrow
K_{\alpha \beta }^{(1)}$ in the second terms, but eq. (6.7)\ has two more
additional terms $K_{1}^{(2)}dX^{1}+2p\rho ^{3}(z)\Gamma _{33}^{r}g_{3r}.$
Therefore, one might guess how the coefficient functions will look like just
by taking into account the above substitution and the contributions from the
additional terms. Revealing the general structure of the coefficient
functions might be particularly useful in higher dimensions, when one would
have a \textquotedblright chain\textquotedblright\ of cubic algebraic
equations. Concretely for the three-dimensional case, investigated here, $%
C_{2}$ in (6.20) can be obtained from $C_{3}$ in (6.4), observing that there
will be an additional contribution from the term $K_{\alpha
}^{(2)}dX^{\alpha }$ for $\alpha =2$. Also, in writing down the coefficient
functions in (6.2)\ it has been accounted that as a result of the previous
parametrization $\frac{a_{3}}{c_{3}}=\rho (z)$ .

Since eq. (6.18)\ is of the same kind as eq. (6.2), for which we already
wrote down the solution, the expression for $dX^{2}$ will be of the same
kind as in formulae (6.17), but with the corresponding functions $%
A_{2},B_{2},C_{2}$ instead of $A_{3},B_{3},C_{3}$. Taking into account (6.19
- 6.20), the solution for $dX^{2}$ can be written as follows: 
\begin{equation}
dX^{2}=\frac{\frac{1}{\sqrt{k_{2}}}\rho (z)\rho ^{^{\prime }}(z)\sqrt{C_{2}}%
+h_{1}(dX^{1})^{2}+h_{2}(dX^{1})+h_{3}}{\frac{1}{\sqrt{k_{2}}}\rho
^{^{\prime }}(z)\sqrt{C_{2}}+l_{1}(dX^{1})^{2}+l_{2}(dX^{1})+l_{3}}\text{ \
\ \ , }  \tag{6.22}
\end{equation}%
where $h_{1},h_{2},h_{3},l_{1},l_{2},l_{3}$ are expressions, depending on $%
\frac{b_{2}}{d_{2}},\frac{d_{2}}{c_{2}},\Gamma _{\alpha \beta }^{r}$ ($%
r=1,2,3$ ; $\alpha ,\beta =1,2$), $g_{\alpha \beta }$, $K_{12}^{(1)}$, $%
K_{21}^{(1)}$ and on the Weierstrass function. 

The representation of the solution for $dX^{2}$ in the form (6.22)\ shows
that it is \textit{an "embedding" solution} of $dX^{1}$ in the sense that it
depends on this function.Correspondingly, the solution (6.17) for $dX^{3}$
is an \textit{"embedding"} one for the variables $dX^{1}$ and $dX^{2}$.

\textbf{Step 4.} It remains now to investigate the one-dimensional cubic
algebraic equation 
\begin{equation}
A_{1}(dX^{1})^{3}+B_{1}(dX^{1})^{2}+C_{1}(dX^{1})+G^{(0)}(g_{ij},\Gamma
_{ij}^{k},R_{ik})\equiv 0\text{ \ \ \ \ ,}  \tag{6.23 }
\end{equation}%
obtained from the two-dimensional cubic algebraic equation (6.18)\ after
applying the linear-fractional transformation 
\begin{equation}
dX^{2}=\frac{a_{2}(z)\widetilde{dX}^{2}+b_{2}(z)}{c_{2}(z)\widetilde{dX}%
^{2}+d_{2}(z)}  \tag{6.24 }
\end{equation}%
and setting up to zero the coefficient function before the highest (third)\
degree of $(dX^{2})^{3}$. Taking into account that as a result of the
previous parametrization $\frac{a_{2}}{c_{2}}=\rho (z)$ , the coefficient
functions $A_{1},B_{1},C_{1}$and $D_{1}$ are given in a form, not depending
on $dX^{2}$ and $dX^{3}$: 
\begin{equation}
A_{1}\equiv 2p\Gamma _{11}^{r}g_{1r}\text{ \ \ \ ,}  \tag{6.25 }
\end{equation}%
\begin{equation}
B_{1}\equiv F_{3}\rho (z)+K_{11}^{(1)}=2p(1+2\frac{d_{2}}{c_{2}})[2\Gamma
_{12}^{r}g_{1r}+\Gamma _{11}^{r}g_{2r}]\rho (z)+K_{11}^{(1)}\text{ \ \ ,} 
\tag{6.26 }
\end{equation}%
\begin{equation*}
C_{1}\equiv F_{1}\rho ^{2}(z)+F_{2}\rho (z)+K_{1}^{(2)}=2p[2\Gamma
_{12}^{r}g_{2r}+\Gamma _{22}^{r}g_{1r}]\rho ^{2}(z)+
\end{equation*}%
\begin{equation}
+(1+2\frac{d_{2}}{c_{2}})(K_{12}^{(1)}+K_{21}^{(1)})\rho (z)+K_{1}^{(2)}%
\text{ \ \ \ \ \ ,}  \tag{6.27 }
\end{equation}%
\begin{equation}
G^{0}\equiv 2p[\Gamma _{22}^{r}g_{2r}+\Gamma _{33}^{r}g_{3r}]\rho
^{3}(z)+K_{22}^{(1)}\rho ^{2}(z)\text{ \ \ \ \ .}  \tag{6.28 }
\end{equation}%
The solution for $dX^{1}$ can again be written in the form (6.17), but with $%
\frac{b_{1}}{c_{1}}$, $\frac{d_{1}}{c_{1}}$, $L_{1}^{(1)}$, $L_{2}^{(1)}$, $%
k_{1}$and $B_{1},C_{1}$ instead of these expressions with the indice
\textquotedblright $3$\textquotedblright .

Taking into account formulaes (6.25 - 6.28)\ for $A_{1},B_{1}$ and $C_{1}$,
the final expression for $dX^{1}$ can be written as 
\begin{equation}
dX^{1}=\frac{\frac{1}{\sqrt{k_{1}}}\rho (z)\rho ^{^{\prime }}(z)\sqrt{%
F_{1}\rho ^{2}+F_{2}\rho (z)+K_{1}^{(2)}}+f_{1}\rho ^{3}+f_{2}\rho
^{2}+f_{3}\rho +f_{4}}{\frac{1}{\sqrt{k_{1}}}\rho ^{^{\prime }}(z)\sqrt{%
F_{1}\rho ^{2}(z)+F_{2}\rho (z)+K_{1}^{(2)}}+\widetilde{g}_{1}\rho ^{2}(z)+%
\widetilde{g}_{2}\rho (z)+\widetilde{g}_{3}}\text{ \ \ \ \ ,}  \tag{6.29}
\end{equation}%
where $F_{1},F_{2},f_{1},f_{2},f_{3},f_{4},\widetilde{g}_{1},\widetilde{g}%
_{2}$ and $\widetilde{g}_{3}$ are functions, depending on $g_{\alpha \beta }$%
, $\Gamma _{\alpha \beta }^{r}$ ($\alpha ,\beta =1,2$) and on the ratios $%
\frac{b_{1}}{c_{1}}$, $\frac{b_{1}}{d_{1}}$, $\frac{b_{2}}{d_{2}}$, $\frac{%
d_{1}}{c_{1}}$, $\frac{d_{2}}{c_{2}}$. 

It is also straightforward to prove that expressions (6.22) for $dX^{2}$ and
(6. 29) for $dX^{1}$do not represent elliptic functions. If one assumes that 
$dX^{1}$ is an elliptic function, then from the standard theory of elliptic
functions [9, 11] it will follow that $dX^{1}$ can be represented as 
\begin{equation}
dX^{1}=K_{1}(\rho )+\rho ^{^{\prime }}(z)K_{2}(\rho )\text{ \ \ \ \ ,} 
\tag{6.30}
\end{equation}%
where $K_{1}(\rho )$ and $K_{2}(\rho )$ depend on the Weierstrass function
only. But this representation will contradict with the expression (6.29)\
for $dX^{1}$- consequently the initial assumption has to be rejected.
Similarly, it can be proved that $dX^{2}$ is not an elliptic function. The
details of this simple proof will be left for the interested reader. 

\section{\ \ COMPLEX \ COORDINATE \ DEPENDENCE \ OF \ THE \ METRIC \ TENSOR
\ COMPONENTS \ FROM \ THE \ UNIFORMIZATION \ OF \ A \ CUBIC \ ALGEBRAIC \
SURFACE}

\bigskip In this section it will be shown that the solutions (6.17), (6. 22)
and (6. 29)  of the cubic algebraic equation (3.5) enable us to express not
only the contravariant metric tensor components through the Weierstrass
function and its derivatives, but the covariant components as well.

Let us write down for convenience the system of equations (6.17), (6. 22)
and (6. 29) for $dX^{1}$, $dX^{2}$ and $dX^{3}$ as ($l=1,2,3$) 
\begin{equation}
dX^{l}(X^{1},X^{2},X^{3})=F_{l}(g_{ij}(\mathbf{X}),\Gamma _{ij}^{k}(\mathbf{X%
}),\rho (z),\rho ^{^{\prime }}(z))=F_{l}(\mathbf{X},z)\text{ \ \ \ ,} 
\tag{7.1}
\end{equation}%
\ where the appearence of the complex coordinate $z$ is a natural
consequence of the uniformization procedure, applied with respect to each
one of the cubic equations from the \textquotedblright
embedded\textquotedblright\ sequence of equations.

Yet how the appearence of the additional complex coordinate $z$ on the
right-hand side of (7.1) can be reconciled with the dependence of the
differentials on the left-hand side only on the generalized coordinates $%
(X^{1},X^{2},X^{3})$ (and on the initial coordinates $x^{1},x^{2},x^{3}$
because of the mapping $X^{i}=X^{i}(x^{1},x^{2},x^{3})$)? The only
reasonable assumption will be that \textit{the initial coordinates depend
also on the complex coordinate}, i.e. 
\begin{equation}
X^{l}\equiv X^{l}(x^{1}(z),x^{2}(z),x^{3}(z))=X^{l}(\mathbf{x,}\text{ }z)%
\text{ \ \ \ \ .}  \tag{7.2}
\end{equation}

\bigskip\ Taking into account the important initial assumptions ($l=1,2,3$) 
\begin{equation}
d^{2}X^{l}=0=dF_{l}(\mathbf{X}(z),z)=\frac{dF_{l}}{dz}dz\text{ \ \ ,} 
\tag{7.3}
\end{equation}%
one easily gets the system of three inhomogeneous linear algebraic equations
with respect to the functions $\frac{\partial X^{1}}{\partial z}$, $\frac{%
\partial X^{2}}{\partial z}$ and $\frac{\partial X^{3}}{\partial z}$ ($%
l=1,2,3$): 
\begin{equation}
\frac{\partial F_{l}}{\partial X^{1}}\frac{\partial X^{1}}{\partial z}+\frac{%
\partial F_{l}}{\partial X^{2}}\frac{\partial X^{2}}{\partial z}+\frac{%
\partial F_{l}}{\partial X^{3}}\frac{\partial X^{3}}{\partial z}+\frac{%
\partial F_{l}}{\partial z}=0\text{ \ \ \ ,}  \tag{7.4}
\end{equation}%
The solution of this algebraic system ($i,k,l=1,2,3$) 
\begin{equation}
\frac{\partial X^{l}}{\partial z}=G_{l}\left( \frac{\partial F_{i}}{\partial
X^{k}}\right) =G_{l}\left( X^{1},X^{2},X^{3},z\right) \text{ \ \ \ \ \ } 
\tag{7.5}
\end{equation}%
represents a system of \textit{three first - order nonlinear differential
equations}.\textbf{\ }A solution of this system can always be found in the
form \textbf{\ } 
\begin{equation}
X^{1}=X^{1}(z)\text{ \ \ ; \ \ \ }X^{2}=X^{2}(z)\text{ \ \ ; \ \ \ \ }%
X^{3}=X^{3}(z)\text{ \ \ \ \ \ \ \ \ \ .}  \tag{7.6}
\end{equation}%
and therefore, the metric tensor components will also depend on the complex
coordinate $z$, i.e. $g_{ij}=g_{ij}(\mathbf{X}(z))$. Note that since the
functions $\frac{\partial F_{i}}{\partial X^{k}}$ in the right-hand side of
(7.5) depend on the Weierstrass function and its derivatives, it might seem
natural to write that the solution of the above system of nonlinear
differential equations $g_{ij}$\ will also depend on the Weierstrass
function and its derivatives 
\begin{equation}
g_{ij}=g_{ij}(X^{1}(\rho (z),\rho ^{^{\prime }}(z),X^{2}(\rho (z),\rho
^{^{\prime }}(z),X^{3}(\rho (z),\rho ^{^{\prime }}(z))=g_{ij}(z)\text{ \ \ \
.}  \tag{7.7}
\end{equation}%
Note however that for the moment we do not have a theorem that the solution
of the system (7.5) will also contain the Weierstrass function\textbf{. }But
in spite of this, the dependence on the complex coordinate $z$ will be
retained.

\section{\protect\bigskip DISCUSSION}

\bigskip\ This paper  continues the investigation of cubic algebraic
equations in gravity theory,  initiated in a previous paper [10].

Unlike in  [10], where the treatment of cubic algebraic equations has been
restricted only to the choice of the contravariant tensor $\widetilde{g}%
^{ij}=dX^{i}dX^{j}$, in the present paper it was demonstrated that under a
more general choice of $\widetilde{g}^{ij}$, there is a wide variety of
algebraic equations of various order, among which an important role play the
cubic equations. Their derivation is based on two important initial
assumptions:

1. The covariant and contravariant metric components are treated
independently, which is a natural approach within the framework of \textit{%
affine geometry} [15 - 18].

2. Under the above assumption, the gravitational Lagrangian (or Ricci
tensor) should remain the same as in the standard gravitational theory with
inverse contravariant metric tensor components.

The proposed approach allows to treat the Einstein's equations as algebraic
equations, and thus to search for separate classes of solutions for the
covariant and contravariant metric tensor components\textbf{.} It can be
supposed also that the existence of such separate classes of solutions might
have some interesting and unexplored until now physical consequences. It has
been shown also that the "transition" to the standard Einsteinian theory \
of gravity can be performed by investigating the intersection with the
corresponding algebraic equations. 

The most important result in this paper is given in Section 6 and is related
to the possibility to find the parametrization functions for a
multicomponent cubic algebraic surface, again by consequent application of
the linear-fractional transformation. The parametrization functions in this
particular case represent complicated irrational expressions of the
Weierstrass function and its derivative, unlike in the standard two -
dimensional case, where they are the Weierstrass function itself and its
first derivative. The advantage of applying the \textit{linear- fractional
transformations} (6.5) and (6.24) is that by adjusting their coefficient
functions (so that the highest - third degree in the transformation equation
will vanish), the following sequence of plane cubic algebraic equations is
fulfilled (the analogue of eq.(65) in [10]): 
\begin{equation}
P_{1}^{(3)}(n_{(3)})m_{(3)}^{3}+P_{2}^{(3)}(n_{(3)})m_{(3)}^{2}+P_{3}^{(3)}(n_{(3)})m_{(3)}+P_{(4)}^{(3)}=0%
\text{ \ \ \ \ \ ,}  \tag{8.1}
\end{equation}%
\begin{equation}
P_{1}^{(2)}(n_{(2)})m_{(2)}^{3}+P_{2}^{(2)}(n_{(2)})m_{(2)}^{2}+P_{3}^{(2)}(n_{(2)})m_{(2)}+P_{(4)}^{(2)}=0%
\text{ \ \ \ \ \ ,}  \tag{8.2}
\end{equation}%
\begin{equation}
P_{1}^{(1)}(n_{(1)})m_{(1)}^{3}+P_{2}^{(1)}(n_{(1)})m_{(1)}^{2}+P_{3}^{(1)}(n_{(1)})m_{(1)}+P_{(4)}^{(1)}=0%
\text{ \ \ \ \ \ ,}  \tag{8.3}
\end{equation}%
where $m_{(3)}$, $m_{(2)}$, $m_{(1)}$ denote the ratios $\frac{a_{3}}{c_{3}}$%
, $\frac{a_{2}}{c_{2}}$, $\frac{a_{1}}{c_{1}}$ in the corresponding linear -
fractional transformations and $n_{(3)}$, $n_{(2)}$, $n_{(1)}$ are the
\textquotedblright new\textquotedblright\ variables $\widetilde{dX}^{3}$, $%
\widetilde{dX}^{2}$, $\widetilde{dX}^{1}$. The \textit{sequence of plane
cubic algebraic equations }(8.1 - 8.3) should be understood as follows: the
first one (8.1) holds if the second one (8.2) is fulfilled; the second one
(8.2) holds if the third one (8.3) is fulfilled. Of course, in the case of $n
$ variables (i. e. $n$ component cubic algebraic equation) the
generalization is straightforward. Further, since each one of the above
plane cubic curves can be transformed to the algebraic equation ($i=1,2,3$) 
\begin{equation}
\widetilde{n}_{(i)}^{2}=\overline{P}_{1}^{(i)}(\widetilde{n}%
_{(i)})m_{(i)}^{3}+\overline{P}_{2}^{(i)}(\widetilde{n}_{(i)})m_{(i)}^{2}+%
\overline{P}_{3}^{(i)}(\widetilde{n}_{(i)})m_{(i)}+\overline{P}_{4}^{(i)}(%
\widetilde{n}_{(i)})  \tag{8.4}
\end{equation}%
and subsequently to its parametrizable form, one obtains the solutions of
the initial multicomponent cubic algebraic equation. 

Finally,  it has been shown that from the expressions (7.1) a system of
first - order nonlinear differential equations can be  obtained, for which
always a solution $X^{1}=X^{1}(z)$, $X^{1}=X^{1}(z)$, $X^{1}=X^{1}(z)$
exists. Thus the dependence on the generalized coordinates $X^{1}$\textbf{, }%
$X^{2}$\textbf{, }$X^{3}$ in the uniformization functions (7.1 ) dissappears
and only the dependence on the complex coordinate $z$ remains, as it should
be for uniformization functions.

\subsection*{\protect\bigskip Acknowledgments}

The author is grateful to Dr. I. B. Pestov (BLTP, JINR, Russia), Dr. D. M.
Mladenov (Theor. Phys.Departm., Fac. of Physics, Sofia Univ., Bulgaria), and
especially to Prof. V. V. Nesterenko (BLTP, JINR, Russia), Dr. O. Santillan
(IAFE, Buenos Aires, Argentina) and Prof. Sawa Manoff (INRNE, BAS, Bulgaria)
for valuable comments, discussions and critical remarks. \ 

This paper is written in memory of \ Prof. S. S. Manoff (1943 - 27.05.2005)
- a specialist in classical gravitational theory. \ 

The author is grateful also to Dr. G. V. Kraniotis (Max Planck Inst.,
Munich, Germany) for sending  me his published paper (ref. [8]).

\end{document}